\documentclass[a4paper]{article}
\usepackage[]{graphicx}
\usepackage{epsfig}
\usepackage{lscape}

\usepackage{longtable}
\usepackage[sectionbib]{natbib}

\begin{document}

\begin{center}
\begin{LARGE}
{\bf Near-infrared absorption properties of oxygen-rich stardust analogues:} \\ 
\vspace{0.35cm}
\end{LARGE}
\begin{Large}
{\bf The influence of colouring metal ions}
\end{Large}
\end{center}










\noindent{\it Manuscript status:} Accepted by {\it Astronomy and Astrophysics} \\

\noindent{\it Authors:} S. Zeidler$^{1}$, Th. Posch$^{2}$, H. Mutschke$^{1}$, H. Richter$^{2}$, O. Wehrhan$^{3}$\\
\noindent$^{1}$Astrophysikalisches Institut und Universit\"atssternwarte, Schillerg\"asschen 2-3, D-07745 Jena, Germany\\
\noindent$^{2}$Institut f\"ur Astronomie, T\"urkenschanzstra{\ss}e 17, A-1180 Wien, Austria\\
\noindent$^{3}$Institut f\"ur Optik und Quantenelektronik, Max-Wien-Platz 1, D-07743 Jena, Germany\\


{\bf Abstract}\\
{Several astrophysically relevant solid oxides and
silicates have extremely small opacities
in the visual and near-infrared in their pure forms. Datasets for the opacities and for the imaginary part $k$ of their complex indices of refraction are hardly 
available in these wavelength ranges.}
{We aimed at determining $k$ for
spinel, rutile, anatase, and olivine, especially
in the near-infrared region. Our measurements were made with impurity-containing, natural, and synthetic stardust analogs.}
{Two experimental methods were used: preparing small sections of
natural minerals and synthesizing melt droplets under the electric
arc furnace. In both cases, the aborption properties of the samples
were measured by transmission spectroscopy.}
{For spinel (MgAl$_2$O$_4$), anatase, rutile (both TiO$_2$), and
olivine ((Mg,Fe)$_{2}$SiO$_{4}$), the optical constants have been
extended to the visual and near-infrared. We highlight that the individual
values of $k$($\lambda$) and the absorption cross section $Q_{abs}$($\lambda$) depend strongly on the content in transition metals like iron.
Based on our measurements, we infer that $k$ values
below 10$^{-5}$ are very rare in natural minerals
including stardust grains, if they occur at all.}
{Data for $k$ and $Q_{abs}$($\lambda$) are important for various
physical properties of stardust grains such as temperature and radiation pressure.
With increasing $Q_{abs}$($\lambda$) due to impurities, the
equilibrium temperature of small grains in circumstellar shells increases
as well. We discuss why and to what extent this is the case.}


\section{Introduction \label{sec:intro}}

Most studies of the absorption properties of stardust analogs are
focused on those wavelength ranges where the respective
minerals or glasses have strong resonance features. This is usually
the case in the ultraviolet range, on the one hand, and
in the mid-infrared (MIR) range, on the other. Between these two
ranges -- i.e.\ between the range of strong electronic
resonances and the range of strong lattice vibration resonances --
there is frequently a lack of data for the complex index of
refraction $m$($\lambda$) = $n$($\lambda$) + i\,$k$($\lambda$).
This is especially true for those insulator materials that are
highly transparent in the visible and near infrared (NIR),
such as magnesium and aluminum silicates and oxides, since their
$k$($\lambda$)-values tend towards values that are too
small to be measured with respect to several standard methods of spectroscopy.
For example, the $k$($\lambda$)-values of pure
MgAl$_{2}$O$_{4}$-spinel would decrease from $10^{-3}$ at about
7$\mu$m {to 2.2 $\times$ $10^{-13}$} at about 2$\mu$m according
to Tropf and Thomas (1991). For $\lambda$ $<$ 2\,$\mu$m, no $k$-data are available at all from this source.

The importance of $k$($\lambda$)-data characterizing stardust in the
visual and NIR region becomes clear by the following
argument. Dust in circumstellar shells mostly absorbs ultraviolet,
visual, and NIR radiation from the stellar radiation field,
thermalizes it via its internal degrees of freedom and re-emits it
at mid-, as well at far-infrared wavelengths. Consider an asymptotic
giant branch (AGB) star with an effective photospheric temperature
of 2500--4000\,K: such an object will emit most of its radiation
energy close to 1\,$\mu$m. Having no information on $k$($\lambda$)
for $\lambda$ $\approx$ 1\,$\mu$m
means nothing else than being unable to calculate the
energy budget of a dust-enriched stellar atmosphere.
Therefore, efforts should be made to close the gaps in
$k$($\lambda$) -- which are usually much wider than eventual gaps in
$n$($\lambda$) -- where they show up in optical databases for
stardust components.

For a very limited number of potential cosmic dust minerals --
namely for spinel, rutile, anatase, and olivine --
this task will be approached in the present paper by
measuring the absorption properties depending on their
content of transition metal ions as 'impurities' (colouring and
absorption-enhancing).

Why do standard optical
databases -- especially those used in astronomical modeling --
often lack absorption data for the transparency regions?
 Reflectance spectroscopy on bulk surfaces, which is very helpful for deriving MIR optical constants,
hardly deliver any reliable constraints on $k$($\lambda$)
in the transparency regime. Here, absorption
coefficients based on reflection measurements are usually nothing
more than the results of extrapolations.
A proper determination of the usually small absorption coefficients
in the near-infrared range require transmission measurements 
at high column densities.
Since scattering losses may compete
with the effects of absorption in the visual and NIR region, homogeneous (bulk) samples are
needed, i.e.\ sections of millimeter or tenths of millimeter
thickness. These are not easy to obtain whether synthetically or
from natural sources.

This paper intends to provide a guideline of how to improve
$k$($\lambda$) data for the task of modeling dust absorption in
the visual and NIR wavelength regions.
The paper is structured in the following way.
Section 2 aims
at characterizing some of the absorption processes relevant to
these regions.
In Section 3, we describe our measurements on
individual stardust analogs, namely spinel, rutile, anatase, and olivine.
We present the spectra and
optical constants derived for these materials, focusing on their
transparency regimes. In the final Section, we discuss a specific astrophysical implication of our work, the influence of the
derived $k$($\lambda$)-values on dust temperatures in a radiative
equilibrium.


\section{The concept of the 'transparency regime' \label{transparency}}

\subsection{Definition and illustration}

\begin{figure}
\flushleft
\includegraphics[width=10cm]{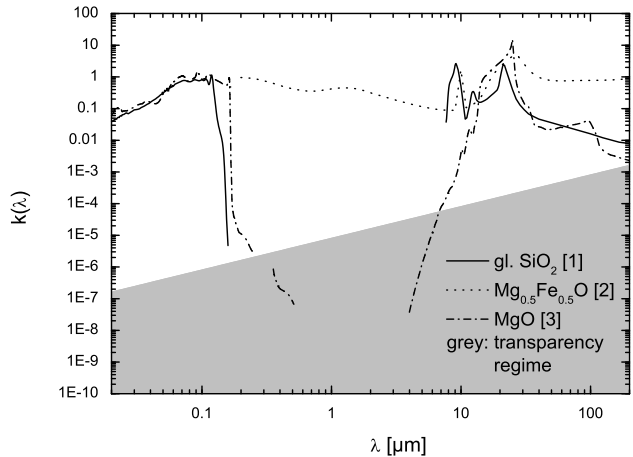}
\caption{Absorption indices $k(\lambda)$ for glassy SiO$_2$,
crystalline Mg$_{0.5}$Fe$_{0.5}$O, and crystalline MgO from the
ultraviolet to the far infrared. The grey area illustrates the
`transparency regime' as defined by Mitra (1985) and by
equation (2). The sources of the data sets are:
[1] Philipp (1985), [2] Henning et al.\ (1995), and
[3] Roessler \& Huffman (1991).} \label{sio2_k}
\end{figure}

Figure 1 shows $k$($\lambda$) for glassy
SiO$_2$ according to Philipp (1985), for
Mg$_{0.5}$Fe$_{0.5}$O according to Henning et al.\ (1995), and for
MgO according to Roessler and Huffman (1991), from the
ultraviolet to the far infrared (FIR). Between the two ranges of
strong absorption, where $k$($\lambda$) reaches and exceeds unity,
a large region in the visual and near-infrared (NIR) can be seen,
where $k$($\lambda$) drops to very low values both for SiO$_2$ and
MgO, while this is not the case for the iron-rich material
Mg$_{0.5}$Fe$_{0.5}$O. Recalling that the absorption coefficient
$\alpha$($\lambda$) of a solid, which enters into Lambert-Beer's
law, is

\begin{equation}
\alpha(\lambda) = \frac{4\pi\,k(\lambda)}{\lambda},
\end{equation}
we can quantitavely {\em define}\/, following Mitra
(1985), the `transparency regime' of a crystalline or
amorphous solid as the region where
\begin{equation}
\alpha(\lambda) < 1\,cm^{-1}. \label{eq:k-criterion}
\end{equation}

At a thickness of 1\,mm, a window in this regime would transmit more
than about 90\% of the light entering at normal incidence; at 1\,cm
thickness, it would still transmit at least 1/$e$ or 37\% of the
radiation (neglecting reflection losses). In terms of
$k$($\lambda$), eq.\ (2) defines a
wavelength-dependent limit as shown in Fig.\,1 (border
between the gray and white areas). At $\lambda$ = 1\,$\mu$m, for example,
we may call a solid `transparent' if $k$\,$<$8\,$\times$\,10$^{-6}$.
This is definitely the case for pure MgO and SiO$_{2}$. It is also
true for pure crystalline MgAl$_{2}$O$_{4}$,
$\alpha$-Al$_{2}$O$_{3}$, Fe-free Mg-silicates and indeed most
oxygen-rich stardust analogs without impurities. We note that for
the visible/near-infrared, our transparency criterion is consistent
with the statement by Bohren \& Huffman (1983) who point out
that already for $k$-values amounting to $k_{vis}$ $\approx$
10$^{-4}$ bulk materials appear {\em black}\/, even in small
pieces.

While the transparency regime is limited towards
shorter wavelengths in the UV/visible by the edge of electronic interband transitions
and possibly excitonic transitions (we give the energies and
wavelengths at which strong absorption sets in Table 1), in the
infrared it is limited by the energetically highest lattice
vibration mode. As a rule of thumb, if a solid has a transparency
region, it sets in at wavelengths shorter than

\begin{equation}
\lambda^{*} = 1/2\,{}\lambda_{LO} = 1 / (2 \omega_{LO}),
\end{equation}
where $\omega_{LO}$ denotes the highest longitudinal optical
frequency (in wavenumbers) and $\lambda_{LO}$ denotes the
corresponding wavelength (cf.\ Barker 1975; however,
Barker uses a different criterion for the transparency and therefore
derives other values for $\lambda$*). Physically, this upper
frequency limit corresponds to the limit up to which two-phonon
absorption processes can be excited. Photons of shorter wavelengths
(i.e.\ in the transparency region) can be absorbed by higher
overtones, i.e.\ the simultaneous excitation of three or more
phonons, which is less likely and does usually not lead to $k$
values above the transparency limit. As can be seen from Table 1,
transparent materials like MgO, amorphous SiO$_{2}$,
MgAl$_{2}$O$_{4}$, and TiO$_{2}$ have two properties in common: their
highest energy LO mode is located at 700 to 1400 cm$^{-1}$, and
accordingly, their transparency regime sets in at wavelengths below
7.1 to 3.6~$\mu$m. This is also the wavelength region where many
sets of optical constants derived from infrared measurements come to
an end because there is a lack of physically meaningful $k$ values (as
mentioned above).

For nonconducting materials, there is an FIR transparency regime as
well. At wavelengths greater than a few hundred micrometers, $k$
again drops to values lower than 10$^{-3}$ to 10$^{-2}$, which are
low enough to fulfill (2). Measurements in this wavelength
range, therefore, bear some similarities to near-infrared
measurements, although wavelengths and absorption mechanisms are
different. Data in this wavelength range are additionally limited by
the availability of FIR-spectrometers.

\subsection{Causes of transparency and of absorption in
the NIR region}

It can furthermore be seen from Fig.\,1 that there are
materials without any transparency regime within the wavelength
range shown. Mg$_{0.5}$Fe$_{0.5}$O -- a solid solution of MgO and
FeO -- is an example of such a material. FeO is a semiconductor, in
contrast to MgO and SiO$_{2}$. Actually, it is a nonstoichiometric 
compound, its sum formula should better be written as Fe$_{1-x}$O. 
Therefore, it contains both Fe$^{2+}$ and Fe$^{3+}$ ions, the 
content of Fe$^{3+}$ is 0.09 according to Henning et al.\ 
(1995). These two properties lead to various
absorption mechanisms for wavelengths in the visible and near
infrared, as well as the FIR, which are not present or occur at
different wavelengths for the previously discussed iron-free oxides,
but are efficient in all the members of the series
Mg$_{x}$Fe$_{1-x}$O (x$<$1). Among these mechanisms are
\begin{enumerate}
\item the fundamental interband absorption edge occurring in the
visible wavelength range (band gap 2.3~eV for FeO, according to 
Park et al.\ 1999);
\item free electron excitations causing the increase of $k$($\lambda$)
towards the FIR. For FeO, this absorption mechanism should
be weak, but a part of the material, which is metastable at room
temperature, may be decomposed into Fe and Fe$_3$O$_4$, which both
have high free-electron densities (at room temperature for
Fe$_3$O$_4$);
\item charge transfer transitions between the orbitals of Fe$^{3+}$
and those of other metal or ligand ions, that are an important source of
strong absorption in the visible extending into the near infrared;
\item the splitting of the d-electron
energy levels in transition metal ions, depending on their coordination by oxygen ions. In
case of lower symmetry, transitions between these d-electron states
can lead to moderately strong absorption bands (so-called crystal
field bands, Burns 1993).
\end{enumerate}

\begin{table}
\centering \caption{Electronic band edges including strong excitons
in terms of energy E$_g$, wavelength $\lambda_g$, longitudial optical lattice 
frequencies $\omega_{LO}$, and onset wavelengths
$\lambda$* for the transparency regions for different materials (see
text for more details).}
\begin{tabular}{p{0.23\linewidth}|p{0.08\linewidth}|p{0.08\linewidth}|p{0.1\linewidth}|p{0.1\linewidth}|p{0.1\linewidth}}
  Material & E$_g$ & $\lambda_g$ & $\omega_{LO}$ & $\lambda$* & Ref.\\
   & [eV] & [$\mu$m] & [cm$^{-1}$] & [$\mu$m] & \\
  \hline
  \hline
  Am.\ SiO$_{2}$ & 8.3 & 0.15 & 1245 & 4.01 & [1]\\
  MgO & 7.5 & 0.17 & 728 & 6.87 & [2]\\
  MgAl$_{2}$O$_{4}$ & 7.75 & 0.16 & 877.2 & 5.7 & [3]\\
  Mg$_{2}$SiO$_{4}$ & 7.5 & 0.17 & 1078 & 4.6 & [4],[5]\\
  TiO$_{2}$ rutile & 2.9 & 0.43 & 831.3 & 6 & [6]\\
  TiO$_{2}$ anatase & 3.2 & 0.39 & 872.8 & 5.7 & [7],[8]\\
  FeO & 2.3 & 0.54 & 526 & 9.5 & [9],[10]\\
  \hline
\end{tabular}
\label{table1}
\flushleft{[1]
Philipp (1985); [2] Roessler \& Huffman (1991); [3]
Tropf and Thomas (1991); [4] Shankland
(1968); [5] Sogawa et al. (2006); [6]
Ribarsky (1985); [7] Tang et al. (1993); [8] Posch
et al. (2003); [9] Park et al. (1999); [10] Henning et al. (1995).}
\end{table}

The last three mechanisms can obviously enhance the absorption in
otherwise transparent wavelength ranges above the transparency
limit. All three of them and even more absorption mechanisms, such
as vibrational excitation, can be introduced by
impurities.

If a (stardust) mineral is transparent at near-infrared and visual
wavelengths in its pure, ideal, stoichiometric form, it may in
reality contain `impurities' that strongly increase $k$($\lambda$).
This phenomenon is well known in geochemistry and mineralogy, but
rarely accounted for in astromineralogical applications. The most
efficient `enhancers' of $k$($\lambda$), consequently of
mass absorption coefficients (opacities), are transition metals:
Sc, Ti, V, Cr, Mn, Fe, Co, and Ni. Among them, Fe, Ni, Cr, and Mn are
probably the most relevant for stardust minerals considering the mean
cosmic elemental abundances (Burns 1993).

\begin{table}
\centering \caption{Mean cosmic elemental abundances of selected
transition metals according to Palme and Beer (1993). \label{t:abund}}
\begin{tabular}{p{0.2\linewidth}|p{0.32\linewidth}}
  Element & abundance log N \\
          & (N(H)=10$^{12}$) \\
  \hline
  \hline
  Sc & 3.1   \\
  Ti & 4.96  \\
  V  & 4.00  \\
  Cr & 5.68  \\
  Mn & 5.45  \\
  Fe & 7.49  \\
  Co & 4.91  \\
  Ni & 6.24  \\
  \hline
\end{tabular}
\end{table}

Any significant inclusion of transition metal ions into an
originally transparent mineral will turn its appearance in the
visual spectral range from colorless to colored or even black. At
present, it is hardly known which amounts of transition metal ions
are actually present in stardust minerals as `impurities', but 
it is unrealistic to expect impurity-free minerals in
circumstellar and interstellar environments.

Even though several visually transparent oxides
such as spinel and Al$_2$O$_3$ have been identified in presolar grains, it has not yet
been possible to determine the transition metal content of these
minerals; in other words, their `color' has not been quantitatively
measured. Spectroscopic methods determining the transition metal
content of cosmic dust by `remote sensing' are restricted, up to
now, to reflectance spectroscopy of lunar and planetary surfaces
(Burns 1993).

Therefore, in the present paper, we chose the following approach
when synthetically including transition metals into oxides: We based
our measurements of $k(\lambda)$ partly on natural terrestrial
samples, which always contain impurities to some extent, and we were partly
able to widen the range of possible inclusions of transition
metals by producing synthetic samples. For one of our samples,
spinel, it was thus possible to create a series of samples and
corresponding sets of optical constants such as exploring the
possible range of absorption that may occur in samples of cosmic
elemental composition.


\section{Individual measured stardust analogs \label{measurements}}

\subsection{Derivation of k from the transmission spectra}

The method we applied to derive $k$($\lambda$) was transmission spectroscopy
of small sections of our materials. For producing these sections we embedded
pieces of our materials with sufficient size (diameter $>$0.5\,mm)
into resin and cut and abraded the resin slabs to thicknesses between 80
and 1100$\mu$m. After that, the samples were just polished to have an
even and smooth surface. 

Some of the examined materials are crystals of non-cubic symmetry and 
therefore optically anisotropic; i.e.\, their
optical constants depend on their orientation relative to the polarization direction of the light.
For these materials, it has been necessary to perform measurements with polarized radiation.
An overview of the sample properties such as chemical composition determined by energy-dispersive 
X-ray analysis (EDX), crystal symmetry, and slab 
thickness is given in Table 3 (the oxygen content is not a direct result of the EDX
measurements, but has been inserted according to stoichiometry).

\begin{table*}[htbp]
\centering
\caption{Compositions of the samples and thicknesses of the platelets
examined in this paper.
\label{t:EDX}}
\begin{tabular}{lccc}
\hline
\textbf{sample} & \textbf{chemical composition} & \textbf{thickness} &
\textbf{crystal symmetry} \\
\hline
\hline
Natural Spinel$^{\dag}$ & Mg$_{1.02}$Al$_{1.93}$Cr$_{0.0087}$Fe$_{0.012}$O$_{4}$ &
250$\mu m$ & cubic \\
Synth.\ Spinel 1 & Mg$_{1.00}$Al$_{1.98}$Cr$_{0.02}$O$_{4}$  & 80$\mu$m & cubic \\
Synth.\ Spinel 2 & Mg$_{0.93}$Al$_{1.99}$Cr$_{0.03}$O$_{4}$  & 270$\mu$m & cubic \\
Synth.\ Spinel 3 & Mg$_{0.98}$Al$_{1.89}$Cr$_{0.12}$O$_{4}$  & 120$\mu$m & cubic \\
Synth.\ Spinel 4 & Mg$_{0.98}$Al$_{1.78}$Cr$_{0.23}$O$_{4}$  & 80$\mu$m & cubic \\
Natural Rutile$^{\dag}$ & Ti$_{0.984}$V$_{0.008}$Fe$_{0.008}$O$_{2}$  & 155$\mu$m & tetragonal \\
Natural Anatase & Ti$_{0.992}$V$_{0.008}$O$_{2}$  & 255 $\mu m$ & tetragonal \\
Olivine San Carlos &  Mg$_{1.96}$Fe$_{0.16}$Si$_{0.89}$O$_{4}$ & 1063/1115$\mu$m & orthorhombic \\
Olivine Sri Lanka &  Mg$_{1.56}$Fe$_{0.4}$Si$_{0.91}$O$_{4}$ & 300$\mu m$ & orthorhombic \\
\hline
\end{tabular}
\flushleft{$\dag$ The natural spinel and rutile samples have small
($<$ 1\%) OH contents that are not shown in the formula.}
\end{table*}

For the spectroscopic measurements, the following spectrometers have been
used: a Perkin Elmer Lambda 19 (for the wavelength range from 0.275 to
2 $\mu$m), a Bruker FTIR 113v (for the wavelength range from 1.6 to 25 $\mu$m),
and a single-beam optical absorption spectrometer constructed on the basis of
a SpectraPro-275 triple-grating monochromator. {The actual range of the transmission 
measurements is limited by the wavelengths at which the samples become opaque. 
For a typical sample thickness of 300\,$\mu$m, this happens for a $k$ value of 
about 10$^{-3}$ on the short-wavelength side and at a few 10$^{-2}$ on the 
long-wavelength side.} 
The transmittance values (transmitted intensity normalized by incident
intensity) were converted in $k(\lambda)$
using the relation

\begin{equation}
k(\lambda)=-\frac{\lambda}{4\pi d} \ln \{\frac{T}{(1-R)^{2}}\},
\label{eq:k}
\end{equation}
where $d$ is the thickness of the small section and the reflectance $R$
is given as
\begin{equation}
R=\frac{(n(\lambda)-1)^2+k(\lambda)^2}{(n(\lambda)+1)^2+k(\lambda)^2}.
\end{equation}
For very small values of $k(\lambda)$ this equation reduces to
\begin{equation}
R=\frac{(n(\lambda)-1)^2}{(n(\lambda)+1)^2}.
\end{equation}
While using $n(\lambda)$ values from Palik (1985) we were able to
derive values for $k(\lambda)$ for various samples.

The following subsections contain a presentation and discussion of
our results for the individual materials.
The derived $k$ values for the NIR will be made publicly available on
the Jena Database of Optical Constants,
http://www.astro.uni-jena.de/Laboratory/Database/databases.html


\subsection{Spinels}

\subsubsection{Material properties, sample preparation and analysis}

Spinel is an abundant component of presolar grains from meteorites
(e.g.\ Hoppe 2004) and has been proposed as the carrier of mid-infrared
emission bands observed in the spectra of circumstellar dust shells (e.g.\ Posch et al.\ 1999, Fabian et al.\ 2001).
Pure stoichiometric spinel (MgAl$_{2}$O$_{4}$) has an elementary
cell that consists of 32\,O anions, 16\,Al, and
8\,Mg cations. In the ideal crystal lattice, the trivalent Al
cations are octahedrally coordinated by oxygen ions, while the
bivalent Mg cations are tetrahedrally coordinated.
The cubic lattice symmetry of this mineral make it optically
isotropic.

While many previous publications refer to impurity-free spinels, the
measurements presented in this paper refer to a series of spinels
that contain impurities (mainly chromium and iron). The Cr
ions partly substitute Al, so they occupy a part of the
octahedral sites. The Fe ions partly substitute Al, partly
Mg. Because the lattice structure of the
Cr-containing spinels is unambiguous (since Cr is not polyvalent in
contrast to Fe), we have chosen synthetic Cr-doped spinels as a
representative of the effect of impurities on $k$, even though in
cosmic environments, it is much more likely to have Fe impurities in
spinels due to the much higher stellar abundance of Fe compared to Cr as
given in Table 2.

Additionally, a dark-red natural spinel crystal from Burma was
examined. Apart from a very small amount of Si, it contains
traces of both Fe and Cr. The exact chemical composition
of the sample is given in Table 3.

\begin{figure}
\flushleft
\includegraphics[width=8.7cm]{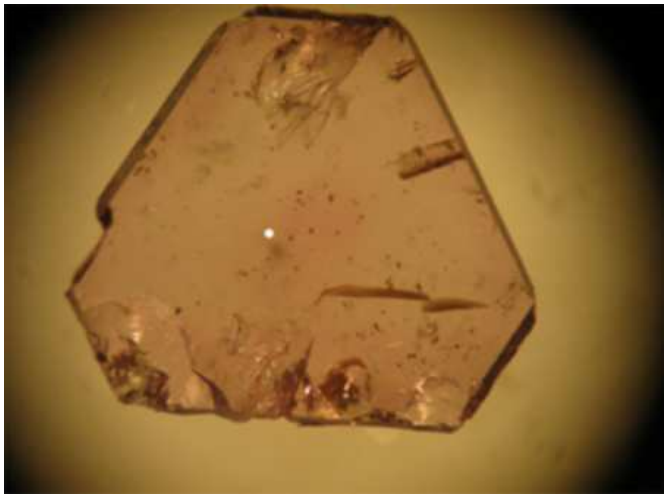}
\caption{Section of the spinel crystal from Burma
used for deriving $k(\lambda)$ from transmission spectroscopy. From
bottom to top this crystal has a size of around 3\,mm.}
\label{f:spinel}
\end{figure}

The synthesis of the Cr-doped spinels was performed
in the following way (see also Richter et al. 2005). We
used a tungsten electric-arc furnace with a water-cooled copper
baseplate. Powders of Mg oxide, Al oxide, and Cr(III) oxide have been
mixed in corresponding proportions so as to obtain the basic
material for the melting procedure. After being homogenized, the
mixture was pressed into several pellets of 150--300\,mg mass each
and a diameter of 1.3\,cm each. This densification of the powder was
necessary to avoid electrostatic interactions between the tungsten
pike in the furnace and the powder particles when starting the
melting process. Since Mg oxide in particular is a hardly fusable
material, the pellets were broken again into several pieces
(4--7 for each pellet) to facilitate the melting process. After
putting several pieces into the furnace, the melting process started
under a 1\,{}bar argon atmosphere. The samples were exposed to the
electric arc until they were fully melted. The formation temperature
should be close to the melting point of spinel (about 2400~K), and it
probably increased with the Cr content (the melting point of
Cr$_2$O$_3$ is about 2700~K). After cooling to subliquidous
temperature within a few seconds and subsequent cooling to room
temperature, we obtained melt-droplets that were 3--5\,mm in size
and had colors ranging from light red in the most Cr-poor
spinel to dark red (almost black) in the most Cr-rich sample. The
theoretical compositions of the samples are given by
MgAl$_{(2-x)}$Cr$_{x}$O$_{4}$, with x amounting to 0.02, 0.03, 0.12,
and 0.23.

The exact chemical compositions were derived from EDX analyses, where several (typically 12) spots for each of
the Cr-spinel droplets were scanned and investigated for
inhomogeneities and variations in the Mg/Al/Cr ratios. Tab.\
3 gives the resulting mean stoichiometries --
also for the TiO$_{2}$ and olivine samples discussed in
Sections 3.3 and 3.4.


\subsubsection{Measured transmission spectra
and derived k values}

\begin{figure}
\flushleft
\includegraphics[width=10cm]{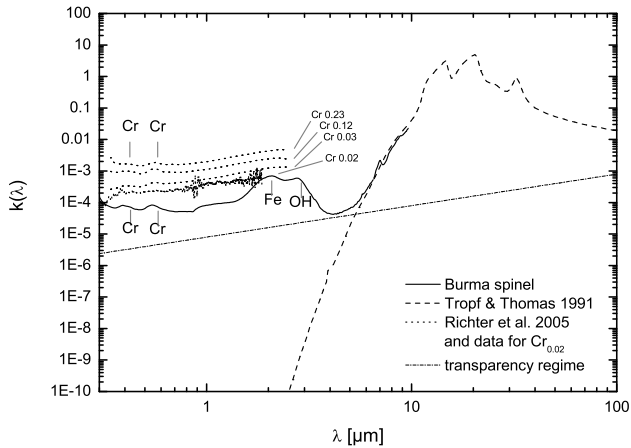}
\caption{Absorption indices $k(\lambda) $ of different spinels derived
from transmission spectroscopy. For the interpretation of the
different bands seen in the spectra, see text.}
\label{f:spinel_k}
\end{figure}

We now present and interpret the results of the transmission
spectroscopy performed on the above-mentioned samples
at UV-, visual, near-infrared, and mid-infrared wavelengths.

The measured transmittances were converted into $k$($\lambda$)
values by applying (4) and Fig. 3
shows the derived results.
The dashed line represents $k$($\lambda$) for pure stoichiometric
spinel according to Tropf and Thomas (1991). The mid
infrared maxima of $k$($\lambda$) -- between 10 and 40$\mu$m --
correspond to three vibrational modes of the lattice structure: The
origin of the 31-32$\mu$m band is an Mg--O vibration, while the remaining
two maxima originate from Al--O vibrations. For further details see Fabian et al.\ (2001).

It can be clearly seen from Fig.\ 3 that for the
impurity-free spinel -- and {\em only}\/ for this one -- a very
steep decrease of $k$($\lambda$) with decreasing wavelength occurs
at $\lambda$ $<$ 7\,$\mu$m.

The solid line denotes $k$($\lambda$) for our natural sample
from Burma which contains about 1\% iron and approximately
as much chromium. This $k$($\lambda$)-curve begins to
deviate from the corresponding line for pure spinel at
6--7\,$\mu$m and never drops below the limit of the
transparency regime, in accordance with the opaque character
of the 0.25mm thick sample. {Within the range of rough 
agreement between the two curves ($>$ 7\,$\mu$m), there 
is some structure in our new data, especially a peak at 
$\sim$7\,$\mu$m, which is likely an overtone of the 13-14\,$\mu$m 
lattice vibration band.}

Between 4 and 1.5\,$\mu$m, a {\em broad}\/ maximum occurs with
its origin in the Fe$^{2+}$ and OH$^{-}$ content of the sample.
Even though the OH content is very small (definitely below
1\%), it is sufficient for the O-H stretching vibration band
to appear in the spectrum as a narrow band at 2.8\,$\mu$m. Then
$k$($\lambda$) approaches values even close to 10$^{-3}$ at
the center of this band. The same holds true for the maximum
of the Fe$^{2+}$ band, which is located at 2.0\,$\mu$m.
A detailed interpretation of the Fe$^{2+}$ and OH$^{-}$ bands
in natural and synthetic spinels can be found in a recent paper
by Lenaz et al.\ (2008). A comparison between our data and
those published by Lenaz et al.\ (2008) is shown in
Fig.\ 4.

\begin{figure}
\flushleft
\includegraphics[width=9.5cm]{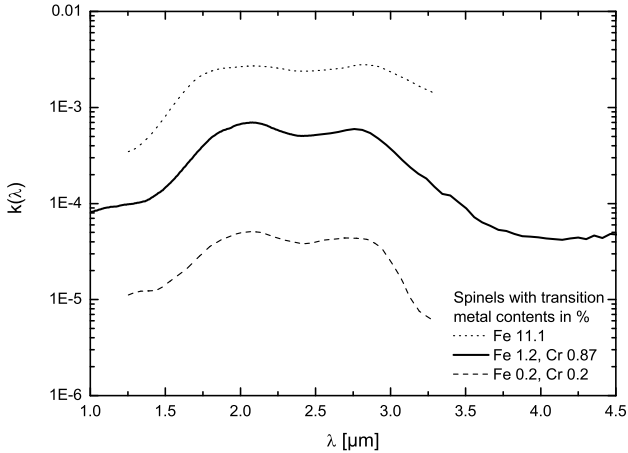}
\caption{Comparison between the absorption indices derived for
spinels with different transtion metal content. The solid line
refers to our sample -- with an Fe content of 1.2\% and a Cr content
of 0.87\%. The dotted line refers to a very Fe-rich sample
(Fe-content of 11.1\%), the dashed line denotes an Fe-poor sample
(both according to Lenaz et al (2008)). The Fe content of the
samples clearly correlates with the respective $k$-values at
2$\mu$m. \label{spinellklenaz}}
\end{figure}

The region below 1\,$\mu$m is characterized by further electronic absorption mechanisms. In this region, the $k$($\lambda$) spectra strongly depend on the Cr content of spinel. This can be nicely seen by comparing the solid line with the dotted lines, which represent the Cr-doped synthetic spinels.

The Al to Cr substitution in the samples results in a generally
increased visual and UV absorption. Additionally, two distinct absorption
bands occur around 0.550\,$\mu$m and 0.387\,$\mu$m. The positions and shapes
of these bands accordance well with the band positions in spectra
of low-Cr Mg-Al- spinels -- both natural and synthetic
ones -- published by Wood et al.\ (1968), Taran et al.\ (1994), and
Ikeda et al.\ (1997).
These bands originate from crystal field electronic transitions,
namely $^4$T$_{2g}$ $ \leftarrow$ $^4$A$_{2g}$ for the 0.550\,$\mu$m maximum
and $^4$T$_{1g}$ $\leftarrow$ $^4$A $_{2g}$ for the 0.387\,$\mu$m
maximum. Of course, the intensity of the Cr electronic absorption bands is higher
for the synthetic Cr-doped samples, and $k$-values up to 2\,$\times$ 10$^{-3}$
are reached at 0.550\,$\mu$m for 12\% Cr content.


\subsection{Rutile and anatase}

Together with Ca and Al oxides, Ti oxides are considered as
candidates for the very first condensates in oxygen-rich circumstellar shells, notwithstanding that titanium is rather rare according to its mean cosmic elemental abundance (2.5 orders of magnitude less abundant than Si and Mg). With respect to dust formation in circumstellar shells, it
is noteworthy that molecular TiO is extremely prominent in the
atmospheres of M- and S-stars. Furthermore, titanium is heavily depleted in the interstellar gas, indicating that it is indeed consumed by previous dust formation processes.

Jeong et al.\ (2003) predict that TiO$_{2}$ is the most promising
candidate for the first (seed-)condensate in oxygen-rich circumstellar shells.
There is also evidence from meteoritics that Ti compounds
form in circumstellar shells and a presolar meteoritic TiO$_{2}$ grain has been
tentatively identified (Nittler 2003).

We examined the two most common terrestrial species of TiO$_{2}$ with
respect to their near infrared absorption properties, namely
anatase and rutile.
{Both materials have a tetragonal crystal symmetry, hence only one optical axis.
Consequently, the measurements for the polarization directions parallel and 
perpendicular to this optical axis could be performed on a single sample cut 
along this axis by rotating the polarizer by 90 degrees.}

$k(\lambda)_{NIR}$ is not unmeasurably small in the case
of our samples due to their Fe and V contents.
The Fe- and V- contents of our TiO$_2$ samples have been determined
by energy-dispersive X-ray measurements and are summarized in
Table 3

\subsubsection{Rutile}

Rutile is the most abundant modification of titanium dioxide on the Earth. 
It has tetragonal symmetry like anatase, but is characterized by a higher density, a greater Ti-Ti distance, and a shorter Ti-O distance than the latter.
As for rutile's ultraviolet and mid-infrared properties, Ribarsky (1985) compiled a rather comprehensive table of optical constants.
However, for wavelengths between 0.4\,$\mu$m and 11.2\,$\mu$m, data
for $k$($\lambda$) are largely missing. 

{We were able to measure $k$ for a natural rutile sample
from 0.5 to 8\,$\mu$m, based on a section with a thickness of 
d=155\,$\mu$m (see Figs.\ 5 and 6).}
The most significant feature in the NIR $k$-spectrum of  
rutile is a sharp peak at 3.03\,$\mu$m. This peak, which becomes
especially prominent when the electric field vector of the polarized
radiation is perpendicular to the c-axis of the crystal, is due to
an OH stretching mode (see Maldener et al.\ 2001). It is
explained by the presence of H$_2$O in natural rutile. The
3.03\,$\mu$m band of our sample may be explained with an H$_2$O content of the order of magnitude of 300 weight ppm, based on the spectra and
analytical results by Maldener et al.\ (2001).

\begin{figure}
\flushleft
\includegraphics[width=8.7cm]{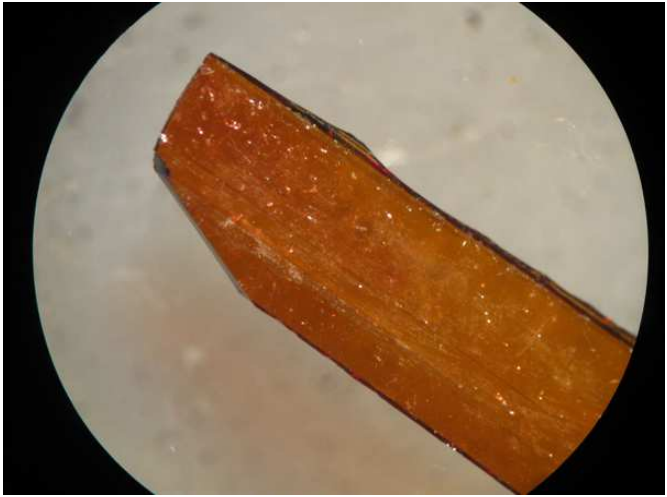}
\caption{Section of the rutile crystal used for
deriving $k(\lambda)$ from transmission spectroscopy. The crystal
has a width of about 1.5\,mm.} \label{rutile}
\end{figure}

\begin{figure}
\flushleft
\includegraphics[width=10cm]{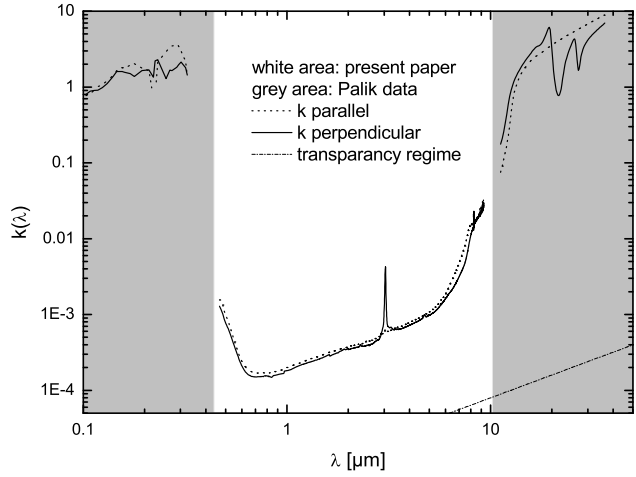}
\caption{Absorption index $k(\lambda)$ for rutile (TiO$_2$)
derived from transmission measurements on a
small section of 155\,$\mu$m diameter.
The $k(\lambda)$ values in the UV and MIR are from
Palik (1985-98).
The sharp peak in $k(\lambda)$ comes from a small crystal water
content of the natural sample.
\label{rutile_k}}
\end{figure}

\subsubsection{Anatase}

Anatase is a modification of tetragonal Ti oxide with a TiO$_6$
octahedron structure consisting of one Ti$^{4+}$ and six O$^{2-}$ ions.
It can be transformed to rutile by heating above 1200\,K.
Its mid-infrared optical constants have been published by
Posch et al.\ (2003).

For anatase we used a section of d=255\,$\mu$m thickness to perform
transmittance measurements {between 0.7 and 8\,$\mu$m}.
{The transmittance spectra obtained for the individual 
orientations of the sample relative to the polarized
radiation did not differ within the measurement's accuracy, 
therefore only one curve is plotted in Fig. 7.}

The most conspicuous feature in the $k(\lambda)$ spectrum
of anatase is the broad double-peaked band 2--3\,$\mu$m.
Maximum $k$ values of more than 0.001 are reached in this range.
The second of the two peaks, located at 3.0\,$\mu$m, corresponds
to an O--H stretching mode at 3389.7\,cm${-1}$. It is an indirect
consequence of the V content of our sample, since V-containing
TiO$_2$ has a stronger hydroxyl content (Zhou et al.\ 2010).
The primary peak in the absorption, located at 2.0\,$\mu$m,
may be directly caused by the V content of the sample.

\begin{figure}
\flushleft
\includegraphics[width=9.5cm]{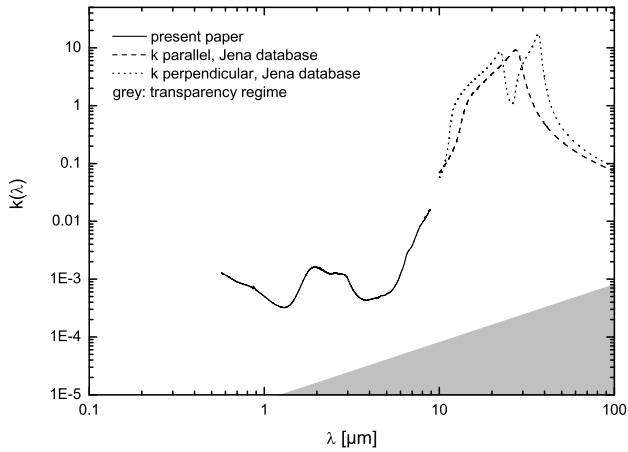}
\caption{Absorption index $k(\lambda)$ of the TiO$_2$ modification
anatase derived from transmission spectroscopy. 
The values for $\lambda>10\,\mu$m are from Posch et al.\ 
(2003).} \label{anatase_k}
\end{figure}


\subsection{Olivine}

Olivine is, together with pyroxene, the most abundant
crystalline silicate in space. Its prominent bands are observed in
many spectra of accretion disks around young stars and in the
outflows of AGB stars (Henning 2003a, 2003b).
Olivine also occurs in our solar system in the form of cometary dust
(Hanner 2003), as interplanetary dust particles (Bradley 2003), and on planetary surfaces (e.g.\ Hartmann 2005).

Olivine has the general sum formula (Mg,Fe)$_{2}$SiO$_{4}$ with its
Mg end member forsterite and the rarer Fe end member fayalite. On
Earth, olivine is the most abundant material in the upper
lithosphere. In general, more Mg-rich and less Fe-rich olivines are
found. It crystallizes in orthorhombic symmetry with isolated
[SiO$_{4}$]$^{4-}$ tedrahedra surrounded by metal cations
(Fe$^{2+}$, Mg$^{2+}$), each in coordination to six oxygen atoms.
There are two different types of these coordinations, both slightly
distorted from the octahedral symmetry. One is centrosymmetric and
elongated along one of the O-O axes (cation sites in this
coordination are called `M1'), and the other is non-centrosymmetric
and irregular (`M2'; cf.\ Burns 1993).

Olivine crystals usually have a greenish to yellowish color, which
can turn into brown when it contains 
traces of Fe$^{3+}$ ions. The greenish coloring is caused by a crystal 
field band of Fe around 1 $\mu$m wavelength, which leads to an increased 
opacity for red light and an increased transmittance of yellow and
green light. The brownish coloring comes from charge transfer
processes toward the Fe$^{3+}$ ions leading to absorption increasing
throughout the visible wavelength range towards the ultraviolet.
This kind of absorption has been detected in the central parts of
our Sri Lanka olivine sample (see below), where cracks may have
given rise to oxidation of Fe$^{2+}$ in interface regions.

In the present paper, we compare near-UV to MIR absorption spectra
of $k$($\lambda$) of San Carlos olivine single crystals with spectra
of an Fe-rich variety of olivine obtained from {Ratnapura}
(Sri Lanka) 
(Fe contents: 8\% and 20\%, respectively -- see Table 3).
Since olivine crystals have non-cubic symmetry, the polarization of
the incoming radiation relative to the three crystallographic axes
$x$, $y$, and $z$ plays an important role in the resulting spectra.
Several platelets of the San Carlos olivine have been prepared to
have a surface perpendicular to one of these axes. Two of them,
denoted as C and D1, have been chosen to represent the properties of
oriented olivine in polarized light due to the quality of their
spectra. Sample C has been cut along the $y$-$z$ plane, sample D1
along the $x$-$y$ plane, so all three crystallographic axes were
available for the measurements. The thicknesses are 1063 $\mu$m for
sample C and 1115 $\mu$m for sample D1. For the Sri Lanka olivine,
only measurements of light polarized along the $y$- and $z$-axes
($E||y$ and $E||z$) were possible, since only one platelet
(thickness: 300 $\mu$m) could be prepared. The sample is shown in
Fig.\ 8. For these measurements we chose an area
of about 1\,mm diameter without cracks, which can be seen in the
figure. This area is also free of brownish colouring which occurs
along the major cracks in the middle of the sample. The surface of
this platelet is also not perfectly oriented perpendicular to the
$x$-axis, because an inclination of approximately 20$^{\circ}$ could be
measured by X-ray diffraction.

The most important differences in the absorption spectra between the
different oriented samples of San Carlos olivine are found in the
bands caused by transitions of the metal cations, located around
0.7--1.5 $\mu$m. While $E||y$ and $E||z$ polarized light creates
broad structures with peaks around 1.056 $\mu$m and 1.108 $\mu$m,
light polarized along the $x$-axis creates a more prominent band
around 1.073 $\mu$m. Shoulders appear in all three polarizations on
both sides of the main peak around 0.87 $\mu$m and between 1.25 and
1.35 $\mu$m. The value of $k$ in the maxima of this band reaches
$\sim$1.8$\times10^{-4}$ for both $E||y$ and $E||z$,
while $k$ reaches a value of $\sim$4$\times10^{-4}$ for $E||x$. The
corresponding values for the Sri Lanka olivine are given in Table
4. The $k$ values of this band are increased
compared to the measurements of the San Carlos olivine in the same
polarization (in its maxima by a factor of $\approx$2) and also the
particular positions of band and shoulders are shifted to longer
wavelengths, except for the short-wavelength shoulder that shifts
to shorter wavelengths.

The continuum $k$ values are in general relatively independent of
the polarization and increases constantly towards longer
wavelengths (see Fig.\ 9). Toward shorter wavelengths we
cannot see an absorption increase, which indicates noticeable charge
transfer processes. The sharp increase in $k$ at around 0.4\,$\mu$m,
which becomes more pronounced with increasing Fe content, is due to
the onset of intrinsic transitions of the Fe$^{2+}$ ions. The
continuum $k$ value is independent of the Fe content of the olivine,
which is remarkable.

Towards longer and towards shorter wavelengths from the crystal
field band at around 1\,$\mu$m, several very weak bands appear. The
bands at shorter wavelengths are comparatively sharp and appear
independently of the polarization at 0.65, 0.49, 0.47, and 0.45
$\mu$m. These bands come from the spin forbidden transitions in the
Fe$^{2+}$ ions (Burns 1993). In contrast to that, the
structures at longer wavelengths show a polarization dependence. For
the Sri Lanka sample in $E||y$ polarization, a broad and relatively
strong band can be seen at 3\,$\mu$m, which is due to the stretching
vibrations of OH groups. For the $E||z$ polarization this band is
much sharper, similar to the corresponding band in the rutile.
Around the 5.5\,$\mu$m wavelength, another band is located, which is likely
an overtone of the 10\,$\mu$m stretching vibration band of the
SiO$_{4}$-tetrahedron.

\begin{figure}
\flushleft
\includegraphics[width=8.7cm]{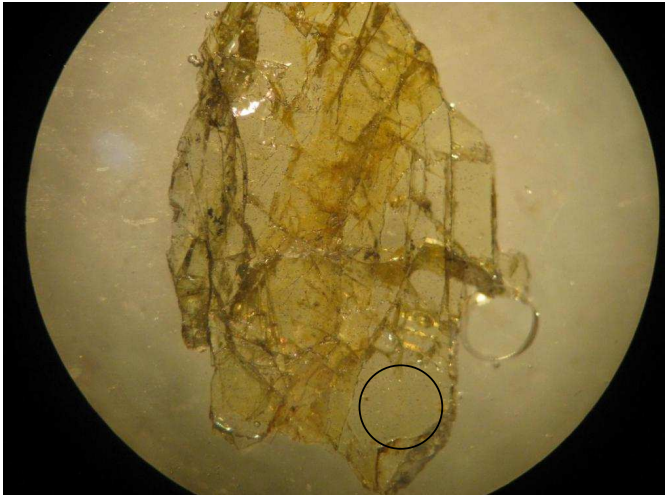}
\caption{Platelet of the Fe-rich Sri Lanka sample used for
transmission measurements under polarized light. The sample has a
width of 4\,mm and a length of 6\,mm. The black circle features the
spot of the measurements.} \label{f:Sri_Lanka}
\end{figure}

\begin{figure}
\flushleft
\includegraphics[width=10cm]{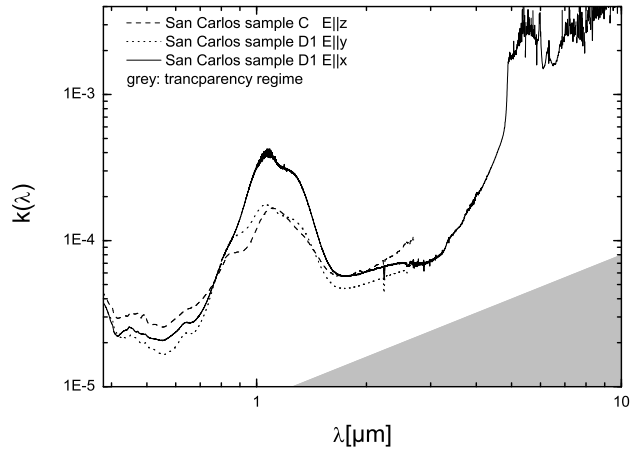}
\caption{Spectra of oriented San Carlos olivine in polarized light.
The spectra were measured from samples C and D1 (see text for further
details).} \label{f:SC_pol}
\end{figure}

\begin{table}
\centering \caption{Specific positions and $k$ values for the main
peaks (`mp') and shoulders (`sh I' and `sh II') of the crystal field band caused by Fe ions of
the San Carlos (SC) and Sri Lanka olivines (SL), respectively.} \label{t:BandsOliv}
\begin{tabular}{l|ll|ll|ll}
  & sh I  & $k_{max}$ & mp & $k_{max}$ & sh II    & $k_{max}$ \\
  & [$\mu$m] & $\times10^{4}$  & [$\mu$m] & $\times10^{4}$ & [$\mu$m] & $\times10^{4}$\\
  \hline
  \hline
  SC    & & & & & &\\
  $E||x$     & 0.827 & 0.90 & 1.074 & 4 & 1.232 & 3.02\\
  $E||y$     & 0.882 & 1.11 & 1.057 & 1.78 & 1.234 & 1.43\\
  $E||z$     & 0.858 & 0.83 & 1.114 & 1.65 & 1.323 & 1.12\\
  \hline
  SL   & & & & & &\\
  $E||y$     & 0.838 & 2.08 & 1.069 & 3.77 & 1.255 & 3.04\\
  $E||z$     & 0.831 & 1.63 & 1.14 & 3.33 & 1.378 & 1.76 \\
  \hline
\end{tabular}
\end{table}

\begin{figure}
\flushleft
\includegraphics[width=10cm]{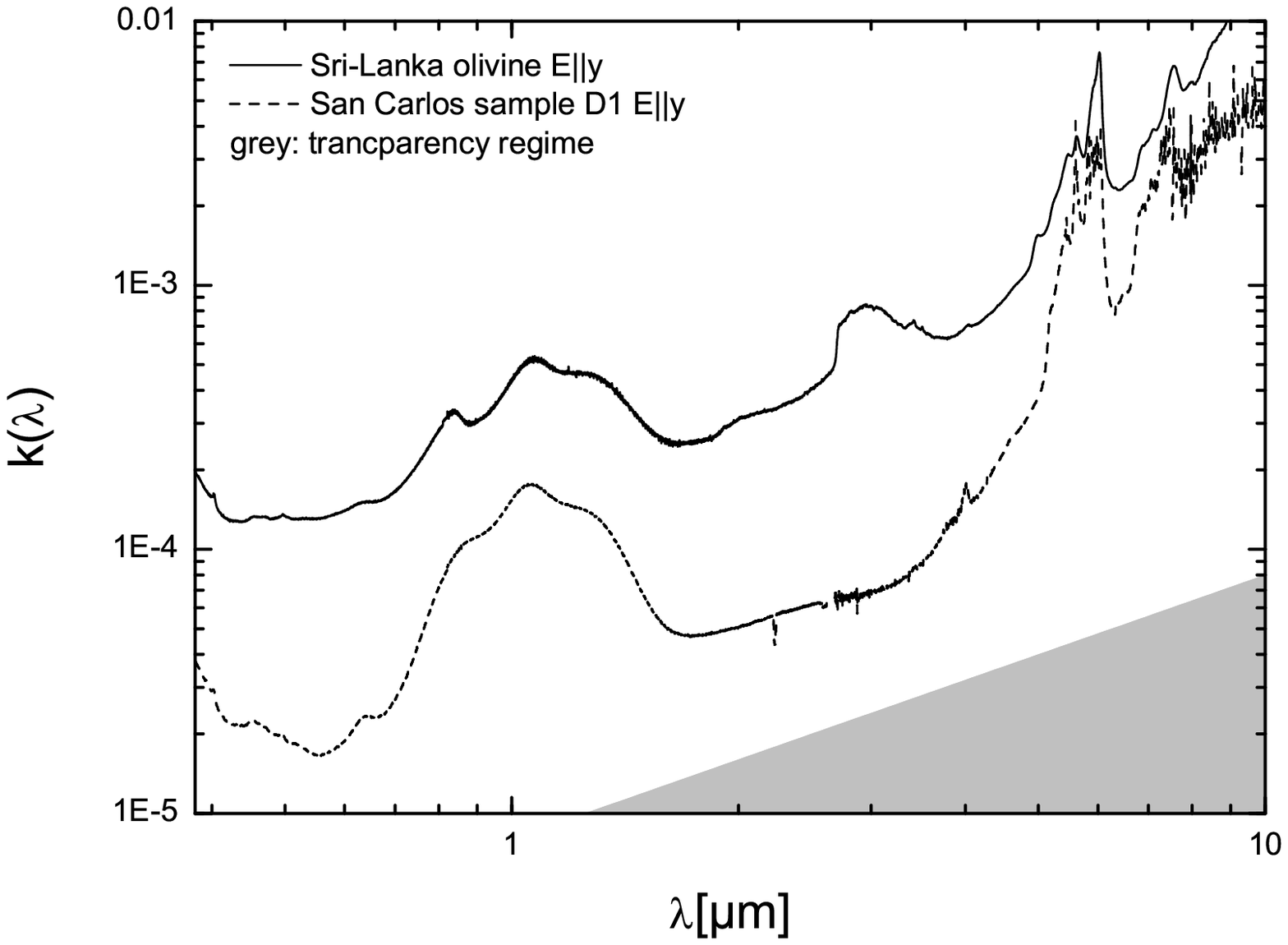}
\caption{Spectra of oriented San Carlos and Sri Lanka olivines in
polarized light for $E||y$. For the San Carlos olivine spectra of
sample D1 are shown.} \label{f:SC-SL_y-pol}
\end{figure}

\begin{figure}
\flushleft
\includegraphics[width=10cm]{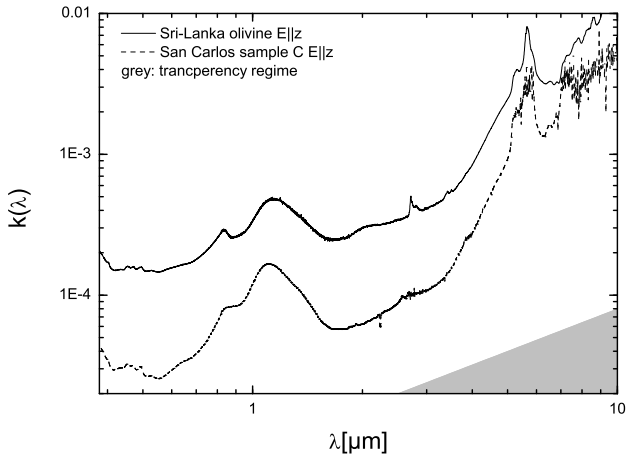}
\caption{Spectra of oriented San Carlos and Sri Lanka olivines in
polarized light for $E||z$. For the San Carlos olivine, spectra of
sample C are shown.} \label{f:SC-SL_z-pol}
\end{figure}


\section{The influence of k on the temperature of small
grains \label{temperature}}

In the present section, we apply our results for the magnitude of
$k$($\lambda$) to the calculation of the radiative equilibrium temperature of dust grains, which is an important parameter characterizing the conditions in dust-forming regions like circumstellar shells, and it indeed
depends on absorption and emission properties.
In the case of a thermodynamical equilibrium between the stellar radiation field 
(assuming a star with radius $R_\mathrm{*}$ with an effective temperature 
$T_\mathrm{*}$) and the dust grains in an optically thin circumstellar shell, 
it is possible to derive the dust temperature $T_\mathrm{d}$ for a given distance from the star from the energy balance.  The radiative energy 
{\em absorbed}\/ by an individual dust grain is then given by

\begin{equation}
{E_\mathrm{abs} = D_\mathrm{rad} \, I_1 },
\end{equation}
where
\begin{equation}
\label{i1}
{I_1 = \int_0^\infty \pi a^2 Q_\mathrm{abs}(\lambda)\pi }
B(\lambda,T_\mathrm{*}) d\lambda,
\end{equation}
and $D_\mathrm{rad}$, the radial dilution factor of the stellar radiation field, is defined as
\begin{equation}
\label{drad}
D_\mathrm{rad} = 2 [1-\sqrt{1-\frac{R_\mathrm{*}^2}{R_\mathrm{d}^2}}].
\end{equation}

Here, $R_\mathrm{*}$ denotes the effective stellar radius, and
$R_\mathrm{d}$ is the distance of a dust grain from the star's
center.
Equation (9) is based on the so-called Lucy-approximation
(Lucy 1971), according to which the star is not considered as a point source, but its spatial extension is taken into account, which is important for small distances of the dust grains from the star.
Already for a distance of 3 stellar radii, the dilution factor $D_\mathrm{rad}$ may be simplified to $R_\mathrm{*}^2$/$R_\mathrm{d}^2$.

The radiative energy {\em emitted}\/ by a single (spherical) dust grain
with the radius $a$ is given by

\begin{equation}
\label{i2}
{E_\mathrm{em} = I_2 
=  \int_0^\infty 4\pi a^2 Q_\mathrm{abs}(\lambda)\pi 
B(\lambda,T_\mathrm{d}) d\lambda. }
\end{equation}
Numerically, the integrals I$_{1}$ and I$_{2}$ cannot (and don't need to) 
be calculated from zero to infinity as equations (8) and (10) suggest. It is sufficient to integrate over those wavelengths where the 
star emits a significant fraction of it total radiative energy.
On the other hand, the energy balance of the dust cannot be calculated correctly
if optical constants (and hence $Q_\mathrm{abs}$-values) are lacking
even at NIR wavelegths where the maximum of the stellar radiation field is
located.

The combination of eqs.\ (8) (10) leads to eq.\
(11), which allows calculating the dependence
of the dust temperature $T_\mathrm{d}$ on the distance from the
star R$_{d}$:

\begin{equation}
\label{rlucy}
R_\mathrm{d}(T_\mathrm{d}) =
\frac{R_\mathrm{*}}{\sqrt{1-[1-\frac{2I_2}{I_1}]^2}}.
\end{equation}
For the simplified dilution factor $R_\mathrm{*}^2$/$R_\mathrm{d}^2$,
the previous equation can also be simplified, namely to

\begin{equation}
R_{\it d}(T_{\it d}) = \sqrt{\frac{I_{1}}{I_{2}}} \, R_{\it *}.
\end{equation}

For very small dust grains (with sizes amounting to 0.01\,$\mu$m or less),
the relation between $T_{\it d}$ and $R_{\it d}$ becomes independent of
the grain size (which enters into the absorption efficiency
$Q_{\it abs}$($\lambda$), in both the integrals I$_{1}$ and
I$_{2}$; see Kr\"ugel 2003). In this case, the influence of the optical
constants on $Q_{\it abs}$($\lambda$) becomes most decisive for the
dust temperature. Again in the small particle limit (also called
Rayleigh limit), the relation between $Q_{\rm abs}$ and the optical
constants $n$ and $k$ is given by

\begin{equation}
\label{ray}
Q_{\rm abs} (\lambda) = 4 \, \frac {2\pi a}{\lambda} \, 
\frac{6 n k}{(n^2-k^2+2)^2 + 4n^2k^2},
\label{eq:Q-nk}
\end{equation}
from which it follows that in those regions where $n$($\lambda$) is
approximately constant, such as in the NIR region for the
oxide species discussed in this paper, $k$($\lambda$) becomes
{\em the} decisive quantity for $Q_{\rm abs}$($\lambda$),
hence decisive for the dust temperature as well.

More precisely, it is the magnitude of $k$($\lambda$)
and $Q_{\rm abs}$($\lambda$) in the visual and NIR {\em relative
to the respective MIR/FIR-values}\/ that determines the dust temperature.
Consequently, dust species that have small $k$($\lambda$), corresponding
to their high transparency in the optical and NIR -- such as transition
metal free oxides and silicates -- will be much less heated by the
stellar radiation than absorbing grain species.
This can be seen most clearly from Fig.\ 12. For MgO, 
the lack of absorption mechanisms operating at short wavelegths leads to inefficient 
radiative heating, such that the dust temperature T$_{d}$ drops below 500\,K already 
at two stellar radii. For Mg$_{0.5}$Fe$_{0.5}$O, on the other hand, T$_{d}$ 
amounts to 2000\,K close to 2\,R$_{*}$. Furthermore, T$_{d}$(R$_{\rm d}$) drops 
much more steeply for MgO than for Mg$_{0.5}$Fe$_{0.5}$O.

The difference between MgO and Mg$_{0.5}$Fe$_{0.5}$O with respect to
NIR absorbance and radiative equilibrium temperature is obviously an
extreme case. Smaller, but still significant differences in
T$_{d}$(r) show up for the other dust species discussed in the
present paper, e.g.\ for spinel (see Fig.\ 13).

As a consequence of the differences in the absorption efficiency
factors calculated according to eq.\ (13) for small
spherical grains -- see insert in  Fig.\ 13 --
a difference in T$_{d}$ results. In this case, it amounts to about
25\% at 2\,R$_{*}$ (the precise values are 850\,K vs.\ 625\,K at
2\,R$_{*}$). While Q$_{abs}$ calculated according to eq.\
(13) for small spherical grains (Fig.\
13 insert) increases linearly with $k$, the increase
in T$_{d}$ in this case appears to be relatively weak, the
gradient of T$_{d}$ with R remains steep for the Cr-containing
spinel.

Notwithstanding, it is important to take the effect of $k_{vis/NIR}$
into account, since when it comes to condensation or evaporation, a
few degrees Kelvin may decide about the existence or nonexistence of
grains in a particular zone around a dust-forming star. Furthermore,
there is still another parameter that linearly increases with $k$ in
the small particle limit, namely the radiation pressure efficiency
factor Q$_{pr}$. For particles that are smaller than the wavelength,
Q$_{pr}$ $\approx$ Q$_{abs}$, and thus an increase in $k$ leads, 
via eq.\ (13), to a corresponding increase in Q$_{pr}$. 
The more efficiently a dust grain absorbs radiation around 1\,$\mu$m, 
the larger the radiation pressure it experiences and the velocity it 
can reach in a radiation-pressure driven stellar wind.
{For a detailed discussion of the influence of absorption {\em and}\/
scattering on the radiation pressure acting on dust grains, we refer 
to H\"ofner (2008).}

\begin{figure}
\flushleft
\includegraphics[width=9.5cm]{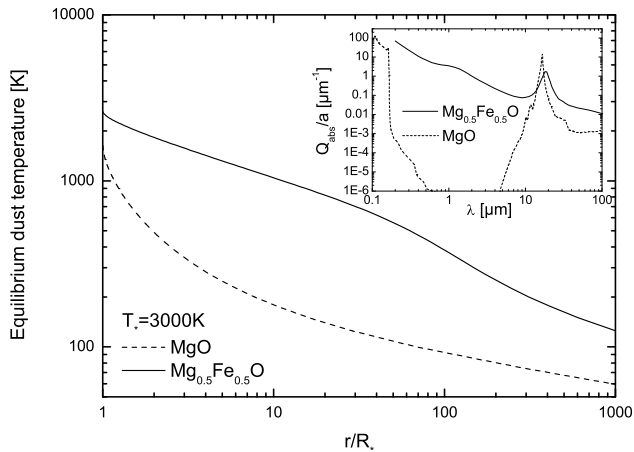}
\caption{Comparison of the radiative equilibrium temperature of MgO
(characterized by extremely low $k$($\lambda$) values in the NIR region)
with the equilibrium temperature of Mg$_{0.5}$Fe$_{0.5}$O
(its $k$($\lambda$) value is more than seven orders of magnitude
larger in the NIR). The stellar effective temperature has been
set to 3000\,K.
\label{f:DustTempMgO} }
\end{figure}

\begin{figure}
\flushleft
\includegraphics[width=9.5cm]{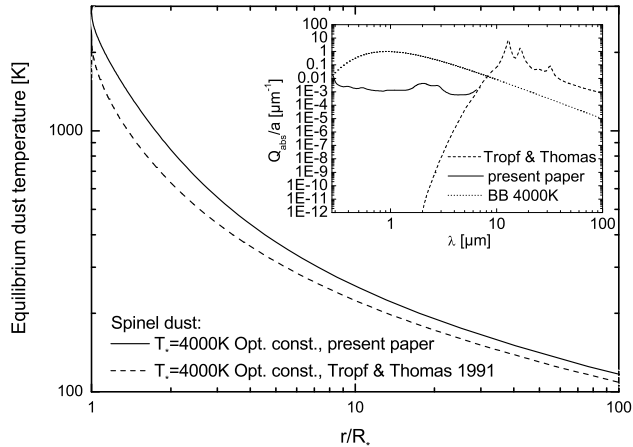}
\caption{The radiative equilibrium temperature of natural (terrestrial)
spinel dust containing impurities compared to the temperature of
impurity-free Mg-Al-spinel according to Palik (1985-98).
The stellar effective temperature has been set to 4000\,K.
\label{f:DustTempSpinel} }
\end{figure}


\section{Conclusions}

We examined the near-infrared and visual absorption properties of
several stardust analogs. Our focus has been on materials for which
extremely low $k$ values -- or no $k$ values at all -- can be
found in the literature for the above-mentioned wavelength range:
spinel, rutile, anatase, and olivine.

The examined samples were either natural terrestrial ones or
synthetic crystals with a defined amount of absorption-enhancing
impurities. In all cases, the impurity contents were such as to be
compatible with cosmic elemental abundances, and in all cases, the
derived $k$ values were 1-2 orders of magnitude higher than the
limit of the `transparency regime' defined at the beginning of this
paper (Eq.\ (1)).

We suggest that even for cosmic dust grains which have smaller impurity contents than our analogs, $k$ values below the limit of transparency will {rarely} be reached. {In cases where no appropriate $k$ data are available,} 
the limit of the `transparency regime' ($k \approx 10^{-5}$ at 1\,$\mu$m and 
$k \approx 10^{-4}$ at 10\,$\mu$m) may serve as a {heuristic} lower limit for the absorption of natural materials including stardust grains, although certain circumstances can produce very transparent natural minerals.

Even though the `real' $k$ values of cosmic
dust grains still has to be determinated by further investigations
(e.g.\ on presolar grains), the above constraint seems more
realistic than adopting $k$ values dropping to 10$^{-8}$ and below,
which can be found in the solid-state physics literature and which mostly
refer to synthetic, impurity-free materials.

As an example, we showed that using $k$ data for spinel taken
from Palik (1985-98) can lead to underestimating the
equilibrium temperature reached by this oxygen-rich dust species in
a circumstellar radiation field. More realistic results can be
obtained if optical constants with $k$ values above the transparency
limit are used. The same holds true for olivine, which is a much more
abundant component of circumstellar dust.

The magnitude of absorption also has an important influence on
other quantities characterizing an expanding circumstellar shell, such as
the radiation pressure. This is, however, beyond the scope of the present
paper.


\begin{small}
\noindent{\em Acknowledgements:}\/
Comments by an anonymous referee helped us to improve the
structure of our paper.
Gabriele Born, Jena, kindly did the sample preparation and 
helped with the EDX measurements. We are grateful to Prof.\ A.\
Tsuchiyama, Osaka University, for providing the Sri Lanka olivine.
HM and SZ acknowledge support by DFG grant Mu 1164/7 within SPP 1385
`The first ten Million Years of the Solar System -- a Planetary
Materials Approach'. TP acknowledges support by the Austrian `Fonds
zur F\"orderung der wissenschaftlichen Forschung' (FWF; project
number P18939-N16). HR is a member of the IK I033-N `Cosmic Matter
Circuit' at the University of Vienna.
\end{small}



\begin{thebibliography}{}

\bibitem[1975]{Barker75}
Barker, A.J., Wilkinson, G.R., Massa, N.E., et al., 1975,
in: Mitra S.S.\ \& Bendow B. (eds.), Optical Properties of Highly
Transparent Solids, Plenum, New York

\bibitem[1983]{BH83}
Bohren, C.F., Huffman D.R., 1983, Absorption and Scattering of Light
by Small Particles, John Wiley, New York

\bibitem[2003]{Bradley03}
Bradley, J., in: Henning Th. (ed.), 2003,
Astromineralogy, Springer-Verlag, Berlin and Heidelberg,
p.\ 225

\bibitem[1993]{Burns93}
Burns, R.G., 1993,
Mineralogical Applications of Crystal Field Theory, 2nd ed.,
Cambridge University Press (= Cambridge Topics in Mineral Physics and
Chemistry, vol.\ 5)

\bibitem[2001]{Fabian01}
Fabian, D., Posch, Th., Mutschke, H., Kerschbaum, F., Dorschner J., 2001,
A\&A 373, 1125

\bibitem[2003]{Hanner03}
Hanner, M.S., 2003, in: Henning Th. (ed.), 2003,
Astromineralogy, Springer-Verlag, Berlin and Heidelberg,
p.\ 173

\bibitem[2005]{Hartmann}
Hartmann, W.K., Moons and Planets, 5th ed., Thomson (Belmont, CA),
2005, p.\ 239

\bibitem[1997]{HeMu97}
Henning, Th. \& Mutschke, H., 1997, A\&A, 327, 743

\bibitem[1995]{Henn95}
Henning, Th., Begemann, B., Mutschke, H., \& Dorschner, J., 1995,
A\&{}AS, 112, 143

\bibitem[1999]{Henn99}
Henning, T., Il'in, V.B., Krivova, N.A., et al., 1999,
A\&{}AS 136, 405

\bibitem[2003]{Henn2003a}
Henning, Th., 2003a, in: Solid-State Astrochemistry, eds.\ V.\ Pirronello, J.\ Krelowski and L.\ Manico, Kluwer, Dordrecht

\bibitem[2003]{Henn2003b}
Henning, Th. (ed.), 2003b, Astromineralogy (= Lecture Notes in Physics,
vol.\ 609), Springer-Verlag, Berlin and Heidelberg

{\bibitem[2008]{Hoef2008}
H\"ofner, S., 2008, A\&A 491, L1}

\bibitem[2004]{Hoppe2004}
Hoppe, P., in: ASP Conf.\ Ser.\ 309, 265

\bibitem[1997]{Ikeda97}
Ikeda, K., Nakamura, Y., Masumoto, K., Shima, H., 1997,
J.\ Amer.\ Cer.\ Soc.\ 80, 2672

\bibitem[2003]{Jeong03}
Jeong, K.S., Winters, J.M., LeBertre, T., Sedlmayr, E., 2003
A\&A 407, 191

\bibitem[2003]{Kruegel03}
Kr\"ugel, E., 2003, The Physics of Interstellar Dust,
Institute of Physics Publishing, Bristol and Philadelphia

\bibitem[2008]{Lenaz08}
Lenaz, D., Skogby, H., Nestola, F., Princivalle, F.,
GCA 72, 475

\bibitem[1971]{L71}
Lucy, L.B., 1971, ApJ, 163, 95

\bibitem[2001]{Mald01}
Maldener, J., Rauch, F., Gavranic, M., Beran, A., 2001,
Mineralogy \& Petrology, 71, 21

\bibitem[1985]{Mitra85}
Mitra, S.S., 1985, in: Palik E.D.\ (ed.), Handbook of Optical
Constants of Solids I, Academic Press, Boston, p.\ 213

\bibitem[2003]{Nittler03}
Nittler, L.\ R. 2003, EPSL 209, 259

\bibitem[1985-98]{Palik85}
Palik, E.D.\ (ed.) 1985--1998, Handbook of Optical Constants
of Solids, 3 vols., Academic Press, Boston

\bibitem[1993]{PB93}
Palme H., Beer H., 1993, in: O. Madelung (ed.), Landolt-B\"ornstein,
Group VI: Astronomy and Astrophysics, Volume 3, p.\ 205

\bibitem[1999]{Park99}
Park, J.-C., Kim, D., Lee, Ch.-S., and Kim, D.-K, 1999,
Bull. Korean Chem. Soc. 20, 1005

\bibitem[1985]{Phil85}
Philipp, H.R., 1985, in: Palik E.\ D.\ (ed.), Handbook of Optical
Constants of Solids I, Academic Press, Boston, p.\ 749

\bibitem[1999]{Posch99}
Posch, Th., Kerschbaum, F., Mutschke, H., et al., 1999,
A\&A 352, 609

\bibitem[1999]{Posch02}
Posch, Th., Kerschbaum, F., Mutschke, H., Dorschner, J.,
J\"ager, C., 2002,
A\&A 393, L7

\bibitem[2003]{Posch03}
Posch, Th., Kerschbaum, F., Fabian, D., et al., 2003,
ApJ Suppl.\ 149, 437

\bibitem[1985]{Rib85}
Ribarsky, M.W., 1985, in: Palik E.\ D.\ (ed.), Handbook of Optical
Constants of Solids I, Academic Press, Boston, p.\ 795

\bibitem[2005]{Richter05}
Richter, H., Posch, Th., Taran, M., Mutschke, H., 2005,
Mineralogy \& Petrology, 85, 53

\bibitem[1991]{RH91}
Roessler, D.M., Huffmann, D.R., 1991, in: Palik E.\ D.\ (ed.),
Handbook of Optical Constants of Solids II, Academic Press, Boston,
919

\bibitem[1968]{Shankland68}
Shankland, T.J., 1968, Science 161, 51

\bibitem[2006]{Sogawa06}
Sogawa, H., Koike, C., Chihara, H., Suto, H., Tachibana, S., et al., 2006,
A\&A 451, 357

\bibitem[1993]{Tang93}
Tang, H., Berger, H., Schmid, P.E., and L\'evy, F., 1993,
Solid State Communications 87, 847

\bibitem[1994]{Taran94}
Taran, M.N., Langer, K., Platonov, A.N., Indutny, V.V., 1994,
Phys.\ Chem.\ Minerals 21, 360

\bibitem[1991]{TroTho91}
Tropf, W.J., Thomas, M.E., 1991,
in: Palik E.\ D.\ (ed.), Handbook of Optical
Constants of Solids II, Academic Press, Boston, p.\ 883

\bibitem[1968]{Wood68}
Wood, D.L., Imbusch, G.F., MacFarlane, R.M., Kisliuk, P., Larkin, P.M., 1968,
J.\ Chem.\ Phys.\ 48, 5255

\bibitem[2010]{Zhou2010}
Zhou, W., Liu, Q., Zhu, Z., Zhang, J., 2010, J.\ Phys.\ D: Appl.\
Phys.\ 43, 035301

\end{thebibliography}
\end{document}